\documentclass[%
prl,
 reprint,
 amsmath,amssymb,
 aps,preprintnumbers]{revtex4-1}

\usepackage[colorlinks=true,linkcolor=black, citecolor=black,
urlcolor=black]{hyperref}

\usepackage{multirow,graphics}
    \newcommand{\beq}{\begin{equation}}
    \newcommand{\eeq}{\end{equation}}
    \newcommand\beqa{\begin{eqnarray}}
    \newcommand\eeqa{\end{eqnarray}}

\renewcommand{\le}{\leqslant}
\renewcommand{\leq}{\leqslant}

\usepackage{amstext}
\usepackage{amssymb}
\usepackage{amsmath}
\usepackage{graphicx}
\usepackage{color}

\begin{document}

\newcommand{\IM}{{\rm Im}\,}
\newcommand{\card}{\#}
\newcommand{\la}[1]{\label{#1}}
\newcommand{\eq}[1]{(\ref{#1})}
\newcommand{\figref}[1]{Fig. \ref{#1}}
\newcommand{\abs}[1]{\left|#1\right|}
\newcommand{\comD}[1]{{\color{red}#1\color{black}}}

\makeatletter
     \@ifundefined{usebibtex}{\newcommand{\ifbibtexelse}[2]{#2}} {\newcommand{\ifbibtexelse}[2]{#1}}
\makeatother

\preprint{ }

\newcommand{\footnoteab}[2]{\ifbibtexelse{%
\footnotetext{#1}%
\footnotetext{#2}%
\cite{Note1,Note2}%
}{%
\newcommand{\textfootnotea}{#1}%
\newcommand{\textfootnoteab}{#2}%
\cite{thefootnotea,thefootnoteab}}}

\def\e{\epsilon}
     \def\bT{{\bf T}}
    \def\bQ{{\bf Q}}
    \def\wT{{\mathbb{T}}}
    \def\wQ{{\mathbb{Q}}}
    \def\ttQ{{\bar Q}}
    \def\tQ{{\tilde \bP}}
        \def\bP{{\bf P}}
    \def\CF{{\cal F}}
    \def\cC{\CF}
     \def\Tr{\text{Tr}}
     \def\l{\lambda}
\def\hbZ{{\widehat{ Z}}}
\def\bZ{{\resizebox{0.28cm}{0.33cm}{\(\hspace{0.03cm}\check {\hspace{-0.03cm}\resizebox{0.14cm}{0.18cm}{\)Z\(}}\)}}}
\newcommand{\rb}{\right)}
\newcommand{\lb}{\left(}

\newcommand{\gT}{T}\newcommand{\gQ}{Q}

\title{Exact Zero Vacuum Energy in twisted \(SU(N)\) Principal Chiral Field}

\author{S\'ebastien Leurent\(^{a}\), Evgeny Sobko\(^{b}\)}

\affiliation{%
\(^{a}\) Universit\'e de Bourgogne Franche-Comt\'e -- Institut de
Math\'ematiques de Bourgogne\\
UMR 5584 du CNRS, 9 avenue Alain Savary 21000 Dijon
\\
 \(^{b}\) DESY Hamburg, Theory Group,
Notkestra{\ss}e 85, 22607 Hamburg, Germany
}

\begin{abstract}
We present a finite set of equations for twisted PCF model. At the special twist in the root of unity we demonstrate that the vacuum energy is exactly zero at any size \(L\). Also in \(SU(2)\) case we numerically calculate the energy of the single particle state with zero rapidity, as a function of \(L\).
\end{abstract}

\maketitle

\section{Introduction}
\(SU(N)\) Principal Chiral Field on a cylinder of circumference \(L\) has the following classical action \(S\):
\begin{gather}
S=\frac{1}{2g^2}\int\limits_{-\infty}^\infty dt \int\limits_{0}^L d x \ \text{tr} \ \partial_\mu U\partial^\mu U^\dag.
\end{gather}

On the quantum level at \(L\rightarrow\infty\) the theory with periodic boundary conditions has \(N-1\) types of
massive excitations which are in one to one correspondence with
fundamental representations of \(SU(N)\). The lowest mass \(m=m_1\) is defined by the mechanism of dimensional transmutation  \(m=m_1=\frac{\Lambda}{g}e^{-\frac{4\pi}{N g^2}}\). All other particles appear as bound states and have masses \(m_k=m\frac{\sin(\frac{\pi k}{N})}{\sin(\frac{\pi}{N})}\). The spectrum of the theory on \(\mathbb{R}^2\) can be completely solved by Bethe Ansatz technique \cite{Wiegmann:1984pw,Ogievetsky:1987vv}. In case of finite \(L\) and periodic boundary conditions the finite system of integral equations with solution in terms of Wronskian determinants was presented in \cite{Kazakov:2010kf}.

In this paper we generalize the construction of \cite{Kazakov:2010kf} to the case of twisted boundary conditions \(U(t,x+L)=e^{i\Phi}U(t,x)e^{-i\Phi}\). In the case of particular twist \(e^{i\Phi}=\Omega=e^{i\frac{\pi}{N}\frac{1+(-1)^N}{2}}\text{diag}[1,e^{\frac{2\pi i}{N}},e^{\frac{4\pi i }{N}},...,e^{\frac{2\pi (N-1) i}{N}}]\) we calculate the energy of single particle state in the rest as a function of \(L\) numerically and show that the vacuum energy is exactly zero. This zero is quite unexpected because PCF is a pure bosonic theory without SUSY.

\section{Twisted TBA}
Using the usual logic of the derivation of TBA we can write the vacuum energy as
\begin{gather}
E_0^d=-\lim \limits_{R\rightarrow \infty} \frac{\log\text{tr} (e^{-H(L)R})}{R}
\end{gather}
and then go to the mirror model \(\text{Tr} (e^{-\tilde{H}(R)L})=\text{Tr} (e^{-H(L)R})\) with mirror Hamiltonian \(\tilde{H}(R)\) and periodic boundary conditions. Twisted boundary conditions \(U(t,x+L)=e^{i\Phi}U(t,x)e^{-i\Phi}\) in the original model can be formalized as an insertion of the defect operator in the mirror model \cite{Ahn:2011xq}:
\begin{gather}
Tr(e^{-\tilde{H}(R)L}e^{i\mathbf{\Phi}})\label{MirrorWithDefectLine}
\end{gather}
where twist \(e^{i\mathbf{\Phi}}\) commutes with the scattering matrix and acts on one particle states as \(e^{i\Phi\otimes 1-1\otimes i \Phi}\) where \(e^{i\Phi}=\text{diag}[e^{i \phi_1},e^{ i \phi_2},...,e^{i \phi_N}]\). This twist leads us to the following TBA equations (\(a\in[1,...,N-1]\)):
\begin{gather}
\mu^{a,0}=m L \frac{\sin\frac{\pi a}{N}}{\sin \frac{\pi}{N}}\cosh\frac{2\pi}{N}u-\log \frac{1}{Y^{a,0}}+\notag\\
+\sum\limits_{a'}\mathcal{K}^{(a',0),(a,0)}\ast \log(1+Y^{a',0})+ \notag\\
 \sum\limits_{s'\neq0,a'}\mathcal{K}^{(a',s'),(a,0)}\ast \log(1+\frac{1}{Y^{a',s'}}),\notag\\
\mu^{a,s\neq0}=-\log Y^{a,s}+\sum\limits_{a'}\mathcal{K}^{(a',0),(a,s)}\ast \log(1+Y^{a',0})+\notag\\
\sum\limits_{s'\neq0,a'}\mathcal{K}^{(a',s'),(a,s)}\ast \log(1+\frac{1}{Y^{a',s'}})\label{TBA}
\end{gather}
with chemical potentials \(\mu^{a,s}\):
\begin{gather}
\begin{cases}
\mu_{a,s}=-is(\phi_{a+1}-\phi_{a}) ,&  s>0\\
\mu_{a,s}=i|s|(\phi_{a+1}-\phi_{a}) ,&  s<0\\
\mu_{a,s}=0,&  s=0
\end{cases}
\end{gather}
and all other ingredients are as in the untwisted case: kernels \(\mathcal{K}\) are derivatives of logarithm of S-matrix which can be read off from \cite{Ogievetsky:1987vv} and Y-functions \(Y^{a,0}=\frac{\rho^{a,0}}{\bar{\rho}^{a,0}}\), \(Y^{a,s\neq 0}=\frac{\bar{\rho}^{a,s}}{\rho^{a,s}}\) are expressed through the  hole and particle's densities \(\bar{\rho}^{a,s}\), \(\rho^{a,s}\). The convolution \(\ast\) is defined in the usual way: \([f\ast g](u)=\int\limits_{-\infty}^\infty dv f(u-v) g(v)\). In terms of Y-functions TBA (\ref{TBA}) can be rewritten as a usual Y-system:
\begin{gather}
\frac{Y_{a,s}^+ Y_{a,s}^-}{Y_{a,s+1}Y_{a,s-1}}=\frac{1+1/Y_{a,s+1}}{1+1/Y_{a+1,s}} \frac{1+1/Y_{a,s-1}}{1+1/Y_{a-1,s}}\label{Y-system}
\end{gather}
At large \(L\) all middle-node Y-functions are exponentially small \(Y_{a,0}=const\times e^{-L p_a}\) and Y-system splits in two independent left and right wings. Here we have introduced the momentum of a's particle \(p_a(\theta)=m\frac{\sin\frac{a \pi}{N}}{\sin\frac{\pi}{N}}\cosh\frac{2\pi}{N}\theta\)  The asymptotic solution at \(L\rightarrow\infty\) can  be written as:
\begin{gather}
Y_{a,s}=\frac{\chi_{a,s+1}(e^{-i\Phi})\chi_{a,s-1}(e^{-i\Phi})}{\chi_{a+1,s}(e^{-i\Phi})\chi_{a-1,s}(e^{-i\Phi})}, \ \ \ s>0 \notag\\
Y_{a,0}=\chi_{a,1}(e^{i\Phi})\chi_{a,1}(e^{-i\Phi})e^{-Lp_a(u)}, \ \ \ s=0 \notag \\
Y_{a,s}=\frac{\chi_{a,|s|+1}(e^{i\Phi})\chi_{a,|s|-1}(e^{i\Phi})}{\chi_{a+1,|s|}(e^{i\Phi})\chi_{a-1,|s|}(e^{i\Phi})}, \ \ \ s<0 \label{LargeLsolution}
\end{gather}
where \(\chi_{a,s}(g)\) is a character of \(g\) in the representation with rectangular Young tableau with \(a\) rows and \(s\) columns.  The ansatz (\ref{LargeLsolution}) obviously solves Y-system (\ref{Y-system}) up to \(O(e^{-mL})\) terms. However Y-system (\ref{Y-system}) doesn't depend on twist and we have to check that solution (\ref{LargeLsolution}) goes through TBA (\ref{TBA}) and reproduces the right chemical potentials. Turn out that a special regularization is needed. Namely we should formally modify the twist as
\begin{gather}
e^{\pm i\Phi}\rightarrow \text{diag}\{e^{\pm i \phi_1}e^{\epsilon_1},e^{ \pm i \phi_2}e^{\epsilon_2},...,e^{\pm i \phi_N}e^{\epsilon_N}\}
\end{gather}
where \(0>\epsilon_1>\epsilon_2>...>\epsilon_N\), \(\frac{\epsilon_i}{\epsilon_{i+1}}\) - fixed and \(\epsilon_N
\rightarrow0\).  Let's show how this prescription reproduces the
chemical potentials in the case of \(a\in[2,...,N-2]\) and
\(s>0\). The corresponding equation reads:
\begin{gather}
\mu_{a,s}=-\log(Y_{a,s}) -\sum\limits_{s'=1}^{\infty} min(s,s')\log(1+\frac{1}{Y_{a-1,s'}})+\notag\\
+\sum\limits_{s'=1}^{\infty} (2 min(s,s')-\delta_{s,s'})\log(1+\frac{1}{Y_{a,s'}})-\notag\\
-\sum\limits_{s'=1}^{\infty} min(s,s')\log(1+\frac{1}{Y_{a+1,s'}})+O(e^{-mL}) \label{a2N-2sneq0}
\end{gather}
Now let's exponentiate its right hand side:
\begin{gather}
e^{\text{r.h.s. of (\ref{a2N-2sneq0})}}=\notag\\
=\lim\limits_{p\rightarrow \infty}\left(\frac{\chi_{a-1,p+1}(e^{i\Phi})}{\chi_{a-1,p}(e^{i\Phi})}\frac{\chi_{a,p}(e^{i\Phi})^2}{\chi_{a,p+1}(e^{i\Phi})^2} \frac{\chi_{a+1,p+1}(e^{i\Phi})}{\chi_{a+1,p}(e^{i\Phi})} \right)^s\label{exprhs}
\end{gather}
Using the above-mentioned epsilon-prescription and the first Weyl formula for characters it's easy to verify that:
\begin{gather}
\lim\limits_{p\rightarrow \infty}\frac{\chi_{a,p+1}}{\chi_{a,p}}= e^{-i\phi_1}e^{-i\phi_2}...e^{-i\phi_a}
\end{gather}
what gives us exponent of l.h.s. of (\ref{a2N-2sneq0}):
\begin{gather}
e^{\text{r.h.s. of (\ref{a2N-2sneq0})}}=\left(\frac{e^{-i \phi_{a+1}}}{e^{-i \phi_{a}}}\right)^s=e^{\mu_{a,s}}
\end{gather}
For other values of \(a\) and \(s\) the consideration is similar.
\section{Exact solution for vacuum at twist $\Omega$}
It is surprising that the asymptotic solution (\ref{LargeLsolution})
turns out to be exact at the twist \(\Omega\). Indeed in case of large
\(L\) middle-node Y-functions \(Y_{a,0}\) was exponentially small,
leading to the decoupling of left and right wings. At the twist \(\Omega\) all characters in fundamental representations are zero \(\chi_{a,0}(\Omega)=0\) and it leads to a vanishing middle-node Y-functions \(Y_{a,0}=0\). This formal solution is singular but it has a natural regularization with above-mentioned epsilon-prescription. In the case of \(\epsilon_i\rightarrow 0\) the ansatz (\ref{LargeLsolution}) solves Y-system (or TBA) up to the terms \(o(\epsilon)\) and gives us solution at any \(L\). Due to the vacuum energy formula:
\begin{gather}
E_{\Omega}^{exact}(L)=\notag\\
=\lim\limits_{\epsilon_i\rightarrow0}-\frac{1}{N}\sum\limits_{a=1}^{N-1}\int\limits_{-\infty}^\infty d\theta p_a(\theta) \log(1+o(\epsilon_i))=0
\end{gather}
we get exact zero for vacuum energy.

At the large \(L\) limit this zero energy can be seen directly from (\ref{MirrorWithDefectLine}). Indeed the leading contribution is a sum over one-particle states and for the particle of the type \(a\) we get factor \(\chi_{a,0}(e^{i \Phi})\chi_{a,0}(e^{-i \Phi})\) which comes from the sum over basis vectors of fundamental representation with \(a\) vertical boxes:
\begin{gather}
\sum\limits_p\langle p|e^{i \mathbf{\Phi}}|p \rangle=\chi_{a,0}(e^{i \Phi})\chi_{a,0}(e^{-i \Phi})
\end{gather}
what gives zero at twist \(\Omega\). The next correction is a L\"{u}scher term corresponding to two-particle contribution and in the \(SU(2)\) case it was calculated in \cite{Ahn:2011xq}. Plugging the twist \(\Omega\) in (3.43) of \cite{Ahn:2011xq} it is easy to verify that the L\"{u}scher term vanishes.

On the other hand, at small \(L\), we have a weakly coupled theory and the one-loop Casimir energy was presented \footnote{At the presence of fermions with \(N_f\) flavors formula (\ref{1loopGenTwist}) modifies \cite{Cherman:2014ofa} as \(E_{e^{i\mathbf{\Phi}}}^{1-loop}=-(N_f-1)\frac{1}{\pi L}\sum\limits_{n=1}^{\infty} \frac{1}{n^2}\left(\left|tr (e^{i n\Phi})\right|^2-1 \right)\). At \(N_f=1\) the theory has \((1,1)\) supersymmetry and the theory has zero vacuum energy.} in \cite{Cherman:2014ofa}:
\begin{gather}
E_{e^{i\mathbf{\Phi}}}^{1-loop}=-\frac{1}{\pi L}\sum\limits_{n=1}^{\infty} \frac{1}{n^2}\left(\left|tr (e^{i n\Phi})\right|^2-1 \right)\label{1loopGenTwist}
\end{gather}
In the periodic case we have the standard Casimir energy
\begin{gather}
E_{periodic}^{1-loop}=-\frac{1}{\pi L}\sum\limits_{n=1}^{\infty} \frac{1}{n^2}(N^2-1)= -\frac{\pi(N^2-1)}{6L},\notag
\end{gather}
corresponding to \(N^2-1\) free bosons \cite{Kazakov:2010kf}.

In the case of twist \(\Omega\) we have:
\begin{gather}
E_{\Omega}^{1-loop}=-\frac{1}{\pi L}\left(\sum\limits_{k=1}^{\infty} \frac{1}{(Nk)^2}(N)^2-\sum\limits_{n=1}^{\infty} \frac{1}{n^2}\right)=0,
\end{gather}
because \(\text{tr} \ \Omega^{n}=0\) for \(n\neq N k\)
\section{ABA}
The Y-system (\ref{Y-system}) can be reformulated in the form of Hirota equations:
\begin{gather}
T^+_{a,s}T^-_{a,s}=T_{a+1,s}T_{a-1,s}+T_{a,s+1}T_{a,s-1}
\end{gather}
and the original Y-functions are expressed through the set of \(\{T_{a,s}(\theta)\}\) as \(Y_{a,s}=\frac{T_{a,s+1}T_{a,s-1}}{T_{a+1,s}T_{a-1,s}}\).

 In the presence of the twist \(e^{i\mathbf{\Phi}}\) the general solution
of Hirota equation 
(up to a gauge degrees of freedom)
has the following generating functional for T-functions \(\lim_{L\to\infty}T^R\):
\begin{gather}
\hat{W}^{R}=\frac{1}{(1-e^{-i\phi_N}X_{(N)}^{R}(\theta) e^{i\partial_\theta})}\cdot\notag\\
\cdot\frac{1}{(1-e^{-i\phi_{N-1}}X_{(N-1)}^{R}(\theta) e^{i\partial_\theta})}\cdot...
\cdot\frac{1}{(1-e^{-i\phi_1}X_{(1)}^{R}(\theta) e^{i\partial_\theta})}\notag\\
=\sum\limits_{s=0}^\infty \frac{T_{1,s}^{R}(\theta+\frac{i}{2}(s-1))}{\varphi(\theta-i\frac{N}{4})}e^{i s \partial_\theta},\label{GenFun}
\end{gather}
where \(X_{k}^{R}\) and \(\varphi(\theta)\) are defined as in
\cite{Kazakov:2010kf}, in terms of some polynomial functions \(Q_{k}^R\)
encoding the excitations defining a given state. The exponent \(R\)
stands for a specific choice of gauge where
\(\lim_{L\to\infty}T_{a,s}^R=0\) is zero if \(s<0\) with
\(0<a<N\), whereas for other values of \((a,s)\) the limit is a
polynomial.

Similarly, there exists a gauge denoted \(T^L\)
where \(\lim_{L\to\infty} T_{a,s}^L\) vanishes if \(s>0\) with \(0<a<N\).
\(T_{1,-s}^L\) has a generating series which differs from the
generating series (\ref{GenFun}) of \(T_{1,s}^R\) by the substitution \(\phi_i\rightarrow-\phi_i\)

Canceling poles at \(T^R_{1,1}\) at \(w-\frac{i}{2}(\frac{N}{2}-k+1)\) we get the twisted auxiliary Bethe equations:
\begin{gather}
-e^{i(\phi_{k-1}-\phi_k)}=\frac{Q^R_{k-2}(w-\frac{i}{2})Q^R_{k-1}(w+i)Q^R_{k}(w-\frac{i}{2})}{Q^R_{k-2}(w+\frac{i}{2})Q^R_{k-1}(w-i)Q^R_{k}(w+\frac{i}{2})}
\end{gather}
and similar for the left wing.

Asymptotic form of middle-node Y-functions reads as:
\begin{gather}
Y_{a,0}(\theta)\sim e^{-m L p_a}\frac{T_{a,1}T^L_{a,-1}}{T_{a+1,0}T_{a-1,0}}\frac{\varphi^{[-\frac{N}{2}-a+1]}\varphi^{[-\frac{N}{2}-a+1]}}{\varphi^{[-\frac{N}{2}+a-1]}\varphi^{[-\frac{N}{2}+a+1]}}\cdot\notag\\
\cdot\frac{1}{\Pi_a\left( (S^{[-\frac{N}{2}]})^2\chi_{CDD}^{[-\frac{N}{2}]} \right)}
\end{gather}
and it leads to the massive Bethe equation \(Y_{1,0}(\theta_j+i\frac{N}{4})=0\) as in the untwisted case\footnote{explicit form of S-matrix \(S\) and CDD factor \(\chi_{CDD}\) can be found in \cite{Kazakov:2010kf}}:
\begin{gather}
-1=\frac{e^{-imL \sinh \frac{2\pi}{N}\theta_j}}{\chi_{CDD}S^2(\theta_j)}\frac{Q^R_{N-1}(\theta_j-\frac{i}{2})Q^L_{N-1}(\theta_j-\frac{i}{2})}
{Q^R_{N-1}(\theta_j+\frac{i}{2})Q^L_{N-1}(\theta_j+\frac{i}{2})}
\end{gather}
\section{Finite L}
Up to some gauge degrees of freedom, the general solution of Hirota equation can be represented through the
following Wronskian determinant:
\begin{gather}\label{eq:3}T_{a,s}^{(R)}(\theta)=
i^{\frac {N(N-1)}2}\mathrm{Det}(c_{j,k})_{1\leq j,k\leq N}\\
  \label{eq:4}
\textrm{where } c_{j,k}=
  \begin{cases}
    e^{i\phi_j(s/2-k)}\overline{q_j}^{[s+a+1+\frac N 2 -2 k]}&\textrm{if
    }k\le a\\
    e^{i\phi_j(-s/2-k)}q_j^{[-s+a+1+\frac N 2 -2 k]}&\textrm{if }k>a\,,
  \end{cases}\nonumber
\end{gather}
in terms of \(2\,N\) complex functions \(\{q_j\,\bar q_j\}\), where we use
the notation \(f^{[n]}(\theta)\equiv f(\theta+n\tfrac i 2)\). It turns
out \cite{Kazakov:2010kf} that \(q_j\) (resp \(\bar q_j\)) is analytic on
the lower (resp upper) half plane, and that
\(q_j\)
and \(\bar q_j\) are complex
conjugated, so that one has \(\overline{T_{a,s}(\theta)}=T_{N-a,s}(\bar{\theta})\). In
addition \(q_j-\bar q_j\) decreases at large \(\theta\) as
\(e^{-L\cosh(2\pi/N\theta)}\), hence the parameterisation
\begin{equation}
  \label{eq:5}
  P_j+i\, \mathcal C*f_j=
  \begin{cases}
    \bar q_j&\textrm{if }\mathrm{Im}(\theta)>0\\
    q_j&\textrm{if }\mathrm{Im}(\theta)<0
  \end{cases}
\end{equation}
 where
\(\mathcal{C}\equiv \frac 1 {2i\pi \theta}\) is the Cauchy Kernel,
\(f_j\equiv i(\bar q_j-q_j)\) is a real jump density and the \(P_j\)'s are
polynomial whose \(L\to\infty\) limit is related (through Wronskian
determinants) to the polynomials \(Q_j^R\). In the periodic case, it was
shown \cite{Kazakov:2010kf} how to write equations for the densities
\(f_j\) for symmetric states, i.e. for states such that
\(T_{a,s}^R=T_{a,-s}^L\).

For the single particle state with zero rapidity, which we denote
\(\Theta_0\) and which defines the mass gap in the periodic case, this
symmetry  \(T_{a,s}^R=T_{a,-s}^L\) is actually broken by the introduction of the
twist. One can however see that for several states such as the vacuum
and this state \(\Theta_0\), the introduction of the
twist preserves a slightly different symmetry:
\begin{equation}
  \label{eq:6}
    {T_{a,-s}^{(L)}(\theta)=(-1)^{\mathcal N} T_{N-a,s}(-\theta)}\,
\end{equation}
where \(\mathcal N\) denotes the number of Bethe roots (for instance it
is zero for vacuum).
One can easily follow the lines of \cite{Kazakov:2010kf} to write
equations for the densities of such state. In what follows we will do
this for both the vacuum and the single-particle state \(\Theta_0\), in
the \(N=2\) case.

For Vacuum we have \( \)\(Q_j^R=1=Q_j^L\) (for all \(j\)), which
corresponds to \(P_j=1\) in \eqref{eq:5}. By contrast, the considered
single-particle state \(\Theta_0\) has \(Q_1^R=Q_1^L=1\)  and
\(Q_2^R(\theta)=Q_2^L(\theta)=\theta\), i.e. \(P_1=1\) and
\(P_2(\theta)=\theta+c\) where \(\lim_{m L\to\infty} c=-\frac 1{2\tan\phi}\).
 Gauge freedom allows to set
\(q_1=1\), and then one can notice that \(T_{1,-1}=-i(\bar q_2-q_2)=-f_2\),
hence the general expression \eqref{eq:3} reduces to
  \begin{multline}\label{eq:tr}
  T_{1,s}^{(R)}=i e^{-i(s+1)\phi}\left((\theta^{[s+1]}+c)^{\mathcal N}+i \mathcal C^{[s+1]}*
    T_{1,-1}^{(R)}\right)\\
-i e^{i(s+1)\phi}\left((\theta^{[-s-1]}+c)^{\mathcal N}+i \mathcal C^{[-s-1]}*
    T_{1,-1}^{(R)}\right)\,,
\end{multline}
where the equation is valid for \(|\mathrm{Im}(\theta)|<s+1\),
we denote \(\phi\equiv\phi_1=-\phi_2\) and the single particle
state 
\(\Theta_0\)
corresponds to
\(\mathcal{N}=1\) while the vacuum 
corresponds to
\(\mathcal N=0\).

If \(\mathcal N=1\), the expression \eqref{eq:tr} involves a constant
\(c\) which is fixed at finite size by the
condition \(\left.T_{1,1}^R\right|_{\theta=0}=0\). This condition
comes from the requirement that \(T_{1,1}^R(\theta)\) and
\(T_{1,1}^L(\theta)=-T_{1,1}^R(-\theta)\) are gauge-equivalent, and if
the gauge transformation is regular this means that the zeroes of
\(T_{1,1}^R\) are symmetric under \(\theta\to-\theta\). Hence the zero
which lies at origin in the \(L\to\infty\) limit stays
exactly at origin at finite \(L\).

If we denote by \(\hat
T_{s}\) the function on the r.h.s. of \eqref{eq:tr}, the equation
  \eqref{eq:tr} now reads: \( |\mathrm{Im}(\theta)|<s+1 \Rightarrow
  \hat T_s (\theta) = T_{1,s}^{(R)} (\theta)\). Then we can follow the
  lines of \cite{Gromov:2008gj} to find integral equations for the function \(T_{1,-1}^{(R)}\) 
entering equations \eqref{eq:tr} 
are
obtained following\cite{Gromov:2008gj}. First, one obtains 
\begin{align}
  \hat T_{0}(\theta)&=
  \begin{cases}
    T_{2,0}^{(R)}(\theta-\tfrac i 2)&\textrm{ if }\mathrm{Im}(\theta)>\tfrac 1 2\\
    T_{1,0}^{(R)}(\theta)&\textrm{ if }\mathrm{Im}(\theta)\in]-\tfrac 1
    2,\tfrac 1 2[\\
    T_{0,0}^{(R)}(\theta+\tfrac i 2)&\textrm{ if }\mathrm{Im}(\theta)<-\tfrac 1 2\,.
  \end{cases}
\end{align}
Next, one gets (for \(|\mathrm{Im}(\theta)|<1/2\))
\begin{align}\label{eq:7}
  Y_{1,0}&= e^{-m L \cosh(\pi\theta)}\frac{(-1)^{\mathcal{N}+\sigma}\hat T_1\{\hat
    T_1\}}{\left(|\hat T_0^{[+2]} \{\hat T_0^{[+2]} \}|^2\right)^{*s}}
  \end{align}
where we use the reality of \(\hat T\) and we denote \(f^{*s}=\exp(\log
f * (1/(2\cosh(\pi \theta))))\).

 The parameter \(\sigma\in\{0,1\}\) is an a
priori remaining freedom, from the fact that the equation for
\(Y_{1,0}^+Y_{1,0}^-\) only fixes \(Y_{1,0}\) up to a sign.
This sign can also be interpreted as an ambiguity in the choice of the branch of the log in
\(f^{*s}=exp(\log(f)*s)\).  In the
periodic case (\(\phi=0\)), this sign is \(\sigma=0\) and it is fixed by
requiring \(Y_{1,0}\) to be a positive function. In our
numeric resolution, we assumed that \(\sigma=0\) holds also at arbitrary twist.

Finally, since we have on the real axis \(Y_{1,0}=\frac{\hat T_1
  T_{1,-1}^{(R)}}{|\hat T_0^{[+1+0]}|^2}\), we get
\begin{equation}\label{eq:8}
  T_{1,-1}^{(R)}=e^{-m L \cosh(\pi\theta)}\frac{(-1)^{\mathcal N+\sigma}|\hat T_0^{[+1+0]}|^2\{\hat
    T_1\}}{\left(|\hat T_0^{[+2]} \{\hat T_0^{[+2]} \}|^2\right)^{*s}}.
\end{equation}
This equation is the twisted analog of equation (43) of
\cite{Gromov:2008gj} for the two states we consider, and it allows to
numerically find their energy \(E=\mathcal{N}-\frac{m}{2} \int \cosh(\pi\theta)\log(1+Y_{10})\mathrm{d}\theta\) 
for arbitrary twist.

The numeric resolution relies on the fact that equations
(\ref{eq:8},~\ref{eq:tr}) give a relation of the form
\(T_{1,-1}^R=F(T_{1,-1}^R)\), where \(F\) is a
contraction mapping whose fixed points are found iteratively.

\section{Numerical Results}
\begin{figure}
  \includegraphics{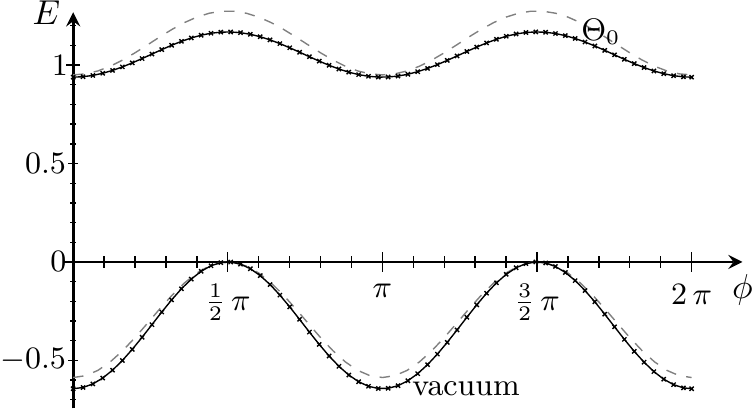}
\caption{Energies of vacuum and \(\Theta_0\) as functions of twist at
  \(L=1\). Dashed gray lines are the large \(L\) expression of energy as a
function of twist. In all numerical calculations we put \(m=1\).}\label{fig:L1}
\end{figure}

\begin{figure}
  \includegraphics{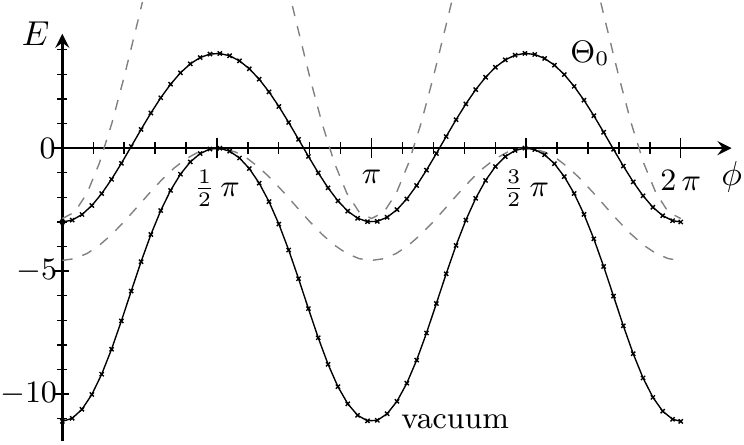}
\caption{Energies of vacuum and \(\Theta_0\) as functions of twist at
  \(L=1/10\). Deviation from the dashed gray lines shows the
  importance of finite-size effects.}\label{fig:L.1}
\end{figure}

At large \(L\), numerical results can be compared to large-L expressions
of the energy (L\"{u}scher corrections). These large L expressions are
obtained from \eqref{eq:7} where the T-functions are replaced with their
large-\(L\) expression, which can be obtained for instance by setting \(T_{1,-1}^R=0\)
in \eqref{eq:tr}. On figures Fig.\ref{fig:L1} and Fig.\ref{fig:L.1} these
large L energies are plotted in gray, while numeric energies of the
vacuum and the state \(\Theta_0\) are plotted in black. We see that the
energy of the states are smooth functions of the twist, which
converge, in the \(\phi\to0\) limit, to
the periodic model's vacuum energy and mass gap, which were already
produced in the literature \cite{Balog:2003yr}.
\begin{figure}
  \includegraphics[scale=0.9]{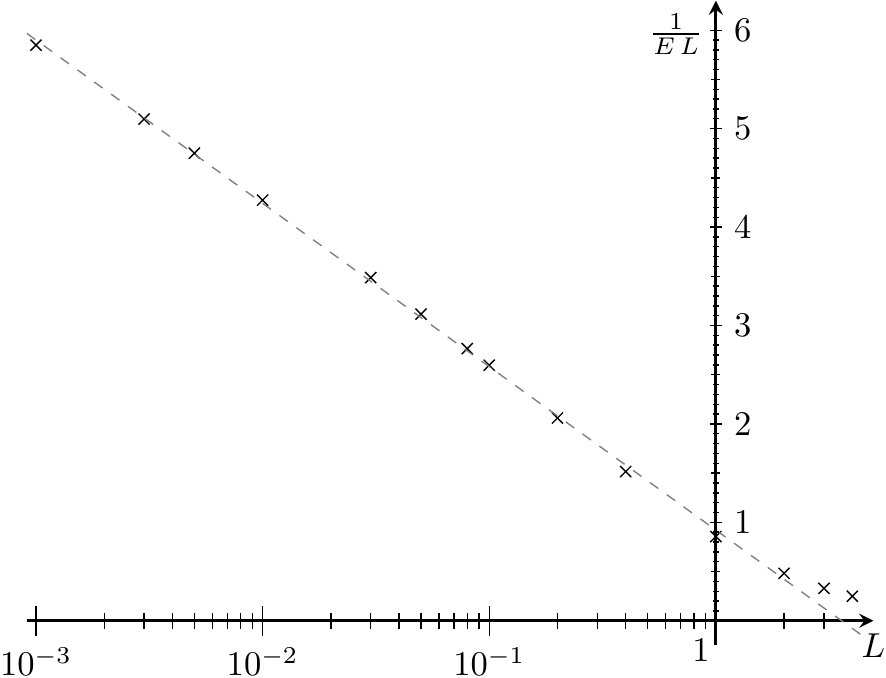}
\caption{Energy of the state \(\Theta_0\) at twist \(\phi=\pi/2\). Crosses
  are numeric results and the dashed gray line is a linear fit. In
  spite of the limited numeric precision, it is manifest that at small
  \(L\), \(\frac 1 {E\, L}\) scales linearly with \(\log L\).}\label{fig:LT0}
\end{figure}

\par\medskip
\section{Discussion}
One of the main results of this paper is the observation that the vacuum energy is exactly zero at the special twist \(\Omega\).
Such behaviour is highly surprising in the absence of supersymmetry which could produce a vanishing vacuum energy by the mechanism of cancellation between fermionic and bosonic degrees of freedom\footnote{The gauge theory with \(N=\infty\) compactified on the small 3d sphere \cite{Basar:2014hda} provides another interesting example of theory with zero vacuum energy but without susy.}. In the language of integrability this vacuum state looks very similar to the vacuum state in deformed AdS/CFT \cite{Ahn:2011xq,Kazakov:2015efa,Arutyunov:2010gu} which has exactly zero energy at the trivial twist, but in \(\mathcal{N}=4\) SYM it is just a consequence of unbroken supersymmetry. It would certainly be fascinating to understand the underlying mechanism or symmetry which is responsible for vanishing of vacuum energy in our case.

In the case of vacuum state the vanishing middle-node Y-functions
\(Y_{a,0}\) can be interpreted as the vanishing of densities \(\rho^{a,0}\) of
massive particles what makes it similar to the vacuum at \(L=\infty\). It is most likely related to the idea of
adiabatic continuity proposed in \cite{Cherman:2013yfa}. Using
resurgence techniques the authors of \cite{Cherman:2013yfa} also proposed
that the mass gap has confined form and scales as \(\sim L\) at small
\(L\). As we see from Fig.\ref{fig:LT0} the state \(\Theta_0\) scales
as in the weak coupling regime and is proportional to \(\frac{1}{L\log L}\). One can
expect that in the presence of the twist, the mass gap is not the difference of
energies of the state \(\Theta_0\) and vacuum. Probably another state
has lower energy than the state \(\Theta_0\), and that either this state
does not correspond to a physical solution of Y-system in the periodic
case, or that it is the same solution as vacuum in this limit. This
would be similar to what happens for \(SU(2)\) Heisenberg spin chains
where the introduction of a twist lifts the degeneracy of some states
belonging to the same multiplet, including states in the multiplet of
vacuum. At the level of the Y-system (or equivalent finite system of integral equations), the search of the
corresponding state is not a trivial task and should be addressed in a
future paper.
\begin{acknowledgments}
\section*{Acknowledgments}
\label{sec:acknowledgments}
We thank D. Dorigoni, V. Kazakov, V. Schomerus and D. Volin for many fruitful and interesting discussions. The work of E.S. was supported by the People Programme (Marie Curie Actions) of the European Union's Seventh Framework Programme FP7/2007-2013/ under REA Grant Agreement No 317089 (GATIS).
S.L. thanks the \emph{centre de calcul de l'université de Bourgogne}
for providing computation ressources.

\end{acknowledgments}


%

\end{document}